# Electronic structure of a realistic model of amorphous graphene.

**V. Kapko[*,1], D. A. Drabold[2] and M. F. Thorpe[1],**

[1] Department of Physics and Astronomy, Arizona State University, Tempe, AZ 85287 USA
[2] Department of Physics and Astronomy, Ohio University, Athens, OH, 45701 USA

[*] Corresponding author: Vitaliy.Kapko@asu.edu

In this note, we calculate the electronic properties of a realistic atomistic model of amorphous graphene. The model contains odd membered rings, particularly five and seven membered rings and no coordination defects. We show that odd-membered rings increase the electronic density of states at the Fermi level relative to crystalline graphene; a honeycomb lattice with semi-metallic character. Some graphene samples contain amorphous regions, which even at small concentrations, may strongly affect many of the exotic properties of crystalline graphene, which arise because of the linear dispersion and semi-metallic character of perfectly crystalline graphene. Estimates are given for the density of states at the Fermi level using a tight-binding model for the $\pi$ states.

## 1 Introduction

Graphene has lately received a great deal of attention as a novel material. Graphene is a two-dimensional network of carbon atoms arrayed in a honeycomb lattice, with the nearest neighbor distance being about 1.42A. The band structure of graphene has been studied by a number of workers. The electronic structure of the material is easy to understand in a tight-binding picture. States near the Fermi level arise from the carbon $p_z$ orbitals perpendicular of the graphene (taken to be in the x-y plane). This central $\pi$ band has a width of order 8.5 eV [1, 2]. States with energies deeper into the valence or conduction bands involve $sp^2$ hybridization.

Since the graphene network is two dimensional, it exhibits some features of basic interest. A primary example is the linear dispersion in the energy bands, which makes graphene electrons analogous to relativistic spin-1/2 particles as treated with the Dirac equation. Consistent with a zero effective mass from the graphene band structure, the experimental carrier mobility is high. Such unusual properties have led to a variety of proposed applications ranging from nano-ribbons to bio-devices.

Most theoretical work on graphene has been carried out for the ideal crystalline material. However, as with most other materials, defects are unavoidable during the preparation of graphene and can play a key role in many observables, and particularly electronic properties.

The consequences of topological and other types of defects on electronic and transport properties of graphene has been studied recently [3]. It has been found that ring disorder is sufficient to introduce gap states. However, these states are localized and their effect on conductivity is limited. In addition, the defects scatter massless Dirac fermions, and that leads to decreasing conductivity [4, 5]. The impact of edges on density of states strongly depends on its type – zigzag or armchair. Only zigzag edges lead to mid gap states, which do not contribute to conductivity as are localized on the edges [3].

The purpose of this Letter is to discuss the electronic properties of amorphous graphene using large and realistic models. We adopt a simple but physically transparent approach to the electronic structure; a tight binding Hamiltonian that is valid near the Fermi level, which is the energy range of primary interest.

## 2 Methods and Models

**2.1 Model of amorphous graphene** was prepared using the Wooten–Weaire–Winer (WWW) method [6, 7]. It was generated by introducing Stone-

Wales defects [8] into perfect honeycomb lattice. The resulting network is presented on figure 1. It contains 800 atoms, each of them three-coordinated, similar to the honeycomb lattice but topologically distinct, with 34.5% of the elementary rings being pentagons, 38% hexagons, 24% heptagons and 4.5% octagons. Since the average size of rings is six, according to Euler's theorem, such a system can exist as a flat 2D structure with some distortions of bond lengths and angles.

Periodic boundary conditions were imposed and the entire network was relaxed with the Keating-like potential [9]:

$$V = \sum_k \left[ \frac{\alpha}{2a^2} \sum_i (\mathbf{r}_{ki}^2 - a^2)^2 + \frac{\beta}{a^2} \sum_{ij} \left( \mathbf{r}_{ki} \cdot \mathbf{r}_{kj} + \frac{a^2}{2} \right)^2 \right] \quad (1)$$

Here $a = 1.42$ Å and $\beta/\alpha = 0.2$. For the structure in figure 1, the root-mean-squared (rms) deviation of the bond length is 5% and rms angle deviation is 16°. Figure 1 also shows distribution of bond lengths. Bonds greater than 1.05 times the average length are in blue and shorter than 0.95 times the average length in red. The remaining bonds are shown in green. The important observation here that blue and red bonds do not alternate as one might have expected, but rather form short chains with blue bonds tending to be associated with large rings (less dense regions), as seen in recent work in other systems [10,11]. One could notice an asymmetry: there are some 6 and 5 long blue chains, but red chains are no longer than 3.

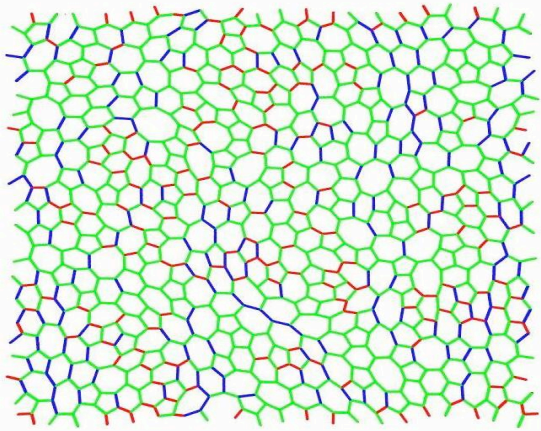

**Figure 1** An 800-atom model of amorphous graphene used in the paper, with periodic boundary conditions. Bonds that are greater than 1.05 times the average length are in blue and less than 0.95 times the average length in red. The remaining bonds are shown in green. Adapted from H. He [7].

The disorder of the network is gauged from the radial distribution function (figure 2), which has typical shape for liquids and amorphous materials. The first peak is very sharp and it corresponds to the bond distance 1.42 Å. Two subsequent peaks can be also distinguished but oscillations of g(r) vanish at distances greater than ~4 Å.

**2.2 Tight-binding approximation** In this work we compare electronic densities of states for crystalline and amorphous graphenes. The first theoretical description of the π and π* electronic bands of crystalline graphene was given by Wallace in 1947 [12]. He developed the tight binding approximation with including nearest- and next-nearest neighbours. The tight-binding Hamiltonian has form

$$H = -\gamma \sum_{\langle i,j \rangle} \left( a_i^+ b_j + H.c. \right) - \gamma' \sum_{\langle\langle i,j \rangle\rangle} \left( a_i^+ a_j + b_i^+ b_j + H.c. \right) \quad (2)$$

where $a_i^+ (a_i)$ are creation (annihilation) operators acting on the A or B sublattice, $\gamma \approx 2.8$ eV is the nearest-neighbor hopping energy and $\gamma' \approx 0.2\gamma$ is the next nearest-neighbor hopping energy. Using Bloch's theorem gives the analytical expression for the band structure:

$$E(\mathbf{k}) = \pm \gamma \sqrt{3 + f(\mathbf{k})} - \gamma' f(\mathbf{k}),$$
$$f(\mathbf{k}) = 2\cos(\sqrt{3} k_y a) + 4\cos\left(\frac{\sqrt{3}}{2} k_y a\right) \cos\left(\frac{3}{2} k_x a\right), \quad (3)$$

where $a$ is the lattice constant, and $\gamma$ and $\gamma'$ are elements of Hamiltonian matrix projected on $2p_z$ atomic wave functions; often treated as fitting parameters. The tight binding approximation for graphene was compared recently with *ab initio* calculation [2], and found to be qualitatively correct, *especially* near the Fermi level of interest here, but to get a good agreement at all energies, at least a third-nearest neighbour approximation is necessary especially because of more the complex hybridization several eV away from the Fermi level.

For a description of topologically disordered solids we will use the tight binding theory [13]. This simplified formulation needs only the information about coordination of nearest neighbours and is valid for both crystalline and amorphous materials. It is based on the Hamiltonian:

$$H = -\gamma \sum_{\langle i,j \rangle} \left( a_i^+ b_j + H.c. \right), \quad (4)$$

which also has only one basis function per atom and only a single parameter $\gamma$ from Eq. 2; and the sum is taken over nearest neighbours. The parameter $\gamma'$ could also be incorporated into this approach, as it is also purely topological, but makes little sense as there would be significant positional dependence based on the local bond angle etc, that would introduce the geometry and add little in terms of understanding close to the Fermi level. The calculation of density of states based on Eq. 4 is simple and reduces the computation of eigenvalues of the connectivity matrix [14]. The Hamiltonian (4) was used in study of electronic structures of amorphous semiconductors [15], who concluded that topological disorder does not eliminate the energy gap, present in the crystalline materials, which is in agreement with experimental data and many subsequent calculations.

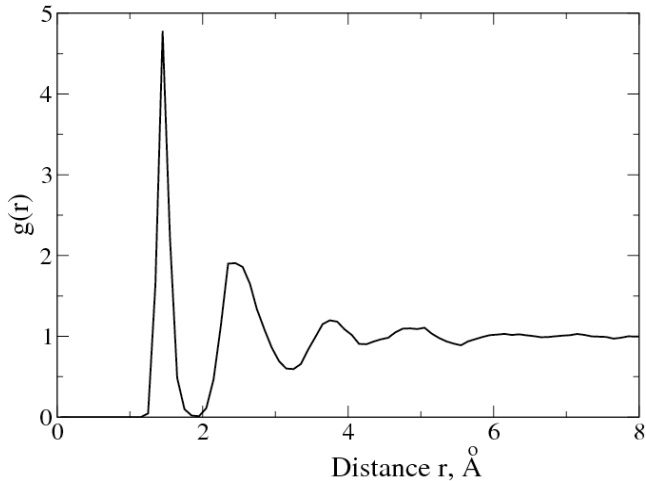

**Figure 2** The radial distribution function $g(r)$ of the model of amorphous graphene, calculated with a bin size 0.1 Å for distances less than 3 Å and 0.2 Å for larger distances.

**3 Results** The main result of the paper, comparison of densities of states of crystalline and amorphous graphene, using Eq. 4 is presented on fig. 3. The solid line, which corresponds to pristine graphene which is semi-metallic, as expected. The black line was calculated for the honeycomb lattice of 800 atoms with periodic boundary conditions, and corresponds to electronic states with zero wavevector in the Brillouin zone of the 800 atom super-cell. The result is almost identical to the infinite lattice calculation (using Eq. 3 with $\gamma' = 0$) denoted by the solid green line, where a full Brillouin zone integration is done using all wavevectors of the super-cell. The ripples on the solid black line are therefore due to finite size effects and serve as an estimate of the error introduced by using an 800 atom super-cell. The role of free boundary conditions is illustrated by blue line. Dangling bonds at the boundary an an 800 atom piece of honeycomb lattice give rise to states near Fermi level. However, their number is proportional to the number of atoms at the boundary and, therefore, vanishes in thermodynamic limit as $1/\sqrt{N}$. The result presented by the dashed red line is calculation for the amorphous system with an 800 atom supercell shown in Fig. 1. In marked and interesting contrast to amorphous bulk semiconductors like silicon, the topological disorder in graphene leads to a significant *increase* of density of states at the Fermi level. Odd rings are even more effective in creating states at the Fermi level than dangling bond defects.

The localization of electronic states can be quantified using the inverse participation ratio, $p$, defined as

$$p^{-1} = \sum_i |\varphi_i|^4, \qquad (4)$$

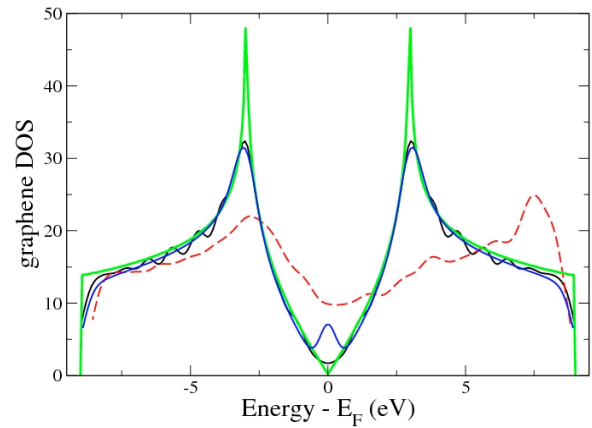

**Figure 3** Density of states of crystalline (solid lines) and amorphous graphene from Fig. 1 (dashed red line). The DOS of the crystal with periodic boundary conditions were solved numerically for an 800 atom super-cell (black line) and by Brillouin zone integration for an infinite lattice (green line), The solid blue line is the DOS of 800 piece on crystalline graphene with free boundaries.

where $\varphi_i$ is a normalized eigenfunction, and $i$ is the atom index. This gives the participation as a function of energy. If the wave function is $1/\sqrt{r}$ on just $r$ atoms and zero everywhere else, then $p = r$ gives an indication of the localization of the wave function. Fig. 4 shows the inverse participation ratios per atom calculated for an 800 atom piece of crystalline graphene (panel a), an 800 atom piece of crystalline graphene with free boundary conditions (panel b) and amorphous graphene (panel c). In the absence of defects all wave functions are highly delocalized with broad distribution of $p/N$ between 0.33 and 0.66. As we have already mentioned, there are no states at the Fermi level for crystalline graphene as shown in panel a. The free boundaries introduce highly localized surface states, which fill in the mid-gap as shown in panel b, and also lead to significant narrowing the distribution $p/N$. Finally, panel c) demonstrates that topological defects cause a significant localization of all the wave functions, and especially those around the Fermi level. These results are consistent with previous studies done on systems with small number of defects [4, 16].

**Conclusions** The electronic structure of an 800 atom model of amorphous graphene with periodic boundary conditions has been studied within the tight-binding approximation, which is realible near the Fermi level. The presence of odd rings in amorphous graphene lead to a very significant increase in the density of states in the vicinity of the Fermi level. It is shown that these states are highly localised. It has recently been shown that such amorphous regions can exist in graphene [17], where they would be expected to have a significant effect in blurring or eliminating some of the exotic effects found in graphene due to the linear dispersion and semi-metallic character of perefectly crystalline graphene.

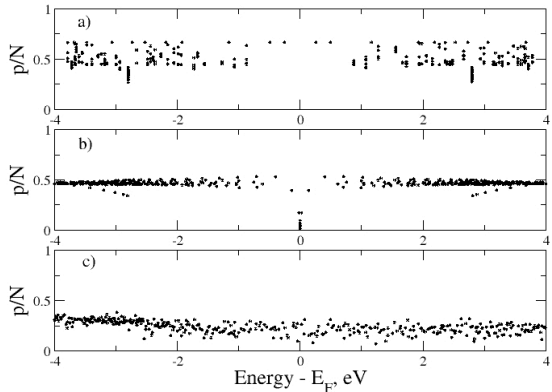

**Figure 4** Inverse participation ratio per atom; all calculated for 800 atom sample. Panel a) for crystalline graphene with periodic conditions, b) for crystalline graphene with free boundaries and c) for amorphous graphene.

**Acknowledgements** We should like to thank Hongxing (Harry) He who first produced the 800 atom model used here for amorphous graphene in 1985 as part of his Ph.D thesis done under the supervision of MFT. We thank Dr. F. Inam for conversations and complementary calculations. VK and MFT thank the National Science Foundation for support under DMR 0703973 and DAD under DMR 0902936.